\documentclass[pdflatex,sn-mathphys-num]{sn-jnl}

\usepackage{graphicx}
\usepackage{multirow}
\usepackage{amsmath,amssymb,amsfonts}
\usepackage{amsthm}
\usepackage{mathrsfs}
\usepackage[title]{appendix}
\usepackage[dvipsnames]{xcolor}
\usepackage{textcomp}
\usepackage{manyfoot}
\usepackage{booktabs}
\usepackage{algorithm}
\usepackage{algorithmicx}
\usepackage{algpseudocode}
\usepackage{listings}
\usepackage{siunitx}
\usepackage{chemformula}
\usepackage{pdfpages}

\begin{document}

\title{Multi-Species Multi-Reaction Model Parametrization of a Commercial NCA Positive Electrode from Half-Cell Open-Circuit Potential Data}

\author*[1]{\fnm{Michele} \sur{Galasso}}\email{galasmic@cvut.cz}

\author[2,3]{\fnm{Petr} \sur{Čech}}

\author[4]{\fnm{Ivo} \sur{Horstkötter}}

\author[5]{\fnm{Simon} \sur{Schwunk}}

\author[2]{\fnm{Zuzana} \sur{Vlčková Živcová}}

\author[1]{\fnm{Václav} \sur{Knap}}

\affil[1]{\orgdiv{Department of Electrotechnology, Faculty of Electrical Engineering}, \orgname{Czech Technical University in Prague}, \orgaddress{\city{Prague}, \country{Czech Republic}}}

\affil[2]{\orgname{Heyrovsk\'{y} Institute of the CAS, v. v. i.}, \orgaddress{\city{Prague}, \country{Czech Republic}}}

\affil[3]{\orgdiv{Department of Physical and Macromolecular Chemistry, Faculty of Science}, \orgname{Charles University}, \orgaddress{\city{Prague}, \country{Czech Republic}}}

\affil[4]{\orgdiv{Chair of Vehicle Mechatronics}, \orgname{Dresden University of Technology}, \orgaddress{\city{Dresden}, \country{Germany}}}

\affil[5]{\orgdiv{Department of Storage System Cell and Control}, \orgname{Dr. Ing. h.c. F. Porsche AG}, \orgaddress{\city{Weissach}, \country{Germany}}}

\abstract{Physics-based models of lithium-ion batteries require the open-circuit potential (OCP) of each electrode as a function of its lithiation. The multi-species multi-reaction (MSMR) model provides a compact, thermodynamically grounded description of this relationship, representing each insertion reaction by a standard potential, a site fraction and a disorder parameter. Here, we parametrize the MSMR model for nickel-cobalt-aluminum oxide (NCA), a cathode chemistry widespread in automotive cells, using the positive electrode of a commercial silicon-graphite/NCA cell. Cathode material was harvested from a pristine cell and from cells cycle-aged to the end of life under two state-of-charge windows, and its half-cell OCP was measured by slow galvanostatic cycling. Smooth differential-capacity curves were obtained by histogram counting, and the charge and discharge branches were fitted with a constrained, scale-invariant optimization that keeps every reaction anchored to a visible peak. We report the first MSMR parameter set for an NCA cathode and show, through a differential-capacity peak analysis across aging states, that its thermodynamic signature is essentially unchanged by aging. This finding supports the common modeling practice of holding the electrode's intrinsic parameters fixed and ascribing aging to electrode-level capacity and alignment changes.}


\begingroup
\renewcommand{\thefootnote}{}
\footnotetext{\textcopyright~2026. This manuscript version is made available under the CC BY-NC-ND 4.0 license \url{https://creativecommons.org/licenses/by-nc-nd/4.0/}. It is the accepted manuscript of an article published in the Journal of Power Sources 700 (2027) 241902, \url{https://doi.org/10.1016/j.jpowsour.2026.241902}.}
\endgroup

\maketitle

\section{Introduction}

Lithium-ion batteries are the energy-storage technology of choice for electric mobility, and their safe and efficient operation depends on a battery management system (BMS) that continuously estimates internal quantities such as the state of charge (SOC), state of health (SOH) and the available energy and power~\cite{Plett2015, Plett2015a}. Almost every state estimation strategy relies, directly or indirectly, on a model of the cell open-circuit voltage (OCV) as a function of SOC, because the OCV ties the measurable terminal voltage to the otherwise unobservable lithiation state of the electrodes~\cite{Plett2015, Lu2021}. Designing OCV models that are simultaneously accurate, physically interpretable, and inexpensive enough to run on embedded hardware is therefore a central problem in battery modeling.

At the cell level, the OCV is the difference between the open-circuit potentials (OCPs) of the positive and negative electrodes, each expressed as a function of its local stoichiometry. Two broad families of OCP representations are used in practice. The first stores the measured potential in a lookup table or fits it with a closed-form empirical expression~\cite{Plett2015}. Such forms can reproduce measured data closely, but they carry no explicit connection to the electrochemical processes inside the electrode, which limits their ability to adapt as the cell ages. The second family is rooted in equilibrium thermodynamics and seeks to describe the OCP from first principles, so that the fitted quantities retain a physical meaning~\cite{Karthikeyan2008, Verbrugge2016, Verbrugge2017}. The latter approach is attractive precisely because aging predominantly alters a small number of physically meaningful quantities (the amount of cyclable lithium and the amount of active material in each electrode), rather than the intrinsic shape of the electrode OCP curves~\cite{Schmitt2021, Dubarry2012}.

Among physics-based descriptions, the multi-species multi-reaction (MSMR) model~\cite{Verbrugge2016, Verbrugge2017} has emerged as a particularly convenient framework. The model treats intercalation as a set of independent reactions, or ``galleries'', each described by three parameters: a standard potential $U^0_j$, a site fraction $X_j$ and a disorder factor $\omega_j$. Beyond its simplicity, the model has the advantage of providing analytically differentiable expressions for both the OCP and the differential capacity, making possible a simple and lightweight software implementation. In the MSMR model, each $U^0_j$ marks a phase transition that appears as a peak in the differential capacity curve, which both aids the diagnosis of electrode behavior and provides a natural way to initialize a parameter fit by inspection~\cite{Verbrugge2017, Lu2021}. This model has been parametrized for a range of active materials, including graphite, silicon, lithium iron phosphate (LFP), spinel manganese oxide (LMO) and layered nickel-manganese-cobalt oxide (NMC)~\cite{Verbrugge2016, Verbrugge2017, Lu2021}.

The interpretability of the MSMR parameters underpins a widely used modeling strategy for aging: the intrinsic gallery parameters \{$U^0_j$, $X_j$, $\omega_j$\} of each electrode are held fixed at their beginning-of-life (BOL) values, and degradation is captured by adjusting only a few cell-level quantities, namely the two electrode capacities and the relative alignment of the electrode stoichiometry windows, from which loss of lithium inventory (LLI) and loss of active material (LAM) at each electrode can be computed~\cite{Dubarry2012}. This separation keeps the aging model simple and well suited to online state estimation, but it relies on the assumption that the shape of each electrode OCP curve is invariant under aging. Recent half-cell measurements support this assumption for nickel-rich cathodes: Schmitt et al. found that the OCP of an NMC-811 positive electrode is essentially unchanged with cycle aging, whereas the change observed in a silicon-graphite negative electrode is dominated by the diminishing capacity contribution of silicon rather than by a change in the underlying potential of either phase~\cite{Schmitt2021}. A reliable BOL parametrization of each electrode is, therefore, the foundation on which such aging-aware models are built.

Two obstacles stand between this attractive framework and its routine use. The first is methodological. Extracting MSMR parameters from half-cell data is not straightforward: low-rate cycling data are corrupted by measurement noise and voltage quantization (the ``data-quality'' problem), constant-current cut-offs can leave part of the curve unmeasured (the ``missing-data'' problem), and practical voltage limits lead to electrode states that are never fully (de)lithiated, so the measured stoichiometry is only relative (the ``inaccessible-lithium'' problem). Lu et al. recently formalized these problems and proposed a systematic processing pipeline to address them~\cite{Lu2021}, while related approaches estimate electrode-level MSMR parameters directly from whole-cell data~\cite{Hu2022}. Nevertheless, the optimization problem for determining MSMR parameters remains prone to non-physical local minima, in which fitted galleries no longer correspond to visible differential-capacity features, or in which the recovered site fractions depart from the model's own conservation constraint $\sum_j X_j = 1$. The site fractions of the NMC-811 set of Garrick et al., for instance, sum to 0.91~\cite{Garrick2024}. A total that departs from unity is a natural outcome of an unconstrained least-squares fit, and is innocuous when the parameters serve as an interpolation of the measured curve; it matters when the $X_j$ are read as site fractions, and, looking at the parameter set alone, it cannot be told apart from a deliberate renormalization of the accessible capacity. Enforcing the constraint during the fit removes this ambiguity. The second obstacle is a simple but important gap in the literature: although MSMR parameter sets have been published for graphite, silicon, LFP, LMO and NMC, to the best of our knowledge no MSMR parametrization has been reported for nickel-cobalt-aluminum oxide (NCA), despite the abundance of NCA cathodes in automotive cells~\cite{Kirst2024}.

In this work, we address both gaps using a commercial silicon-graphite/NCA cell, the Molicel INR-18650-M35A, whose cathode chemistry is identified as NCA by the cell's UN 38.3 transport-test report~\cite{BVCPS2019}. We harvested cathode material from a pristine cell and from cells cycle-aged to the end of life (EOL) under two different SOC windows, we assembled coin cells against lithium metal and measured the cathode OCP by slow cycling. Following the pipeline of Lu et al., we use the histogram-counting method to obtain smooth differential-capacity curves and the separate-branch MSMR fit (their ``Method 5'') to recover the absolute-stoichiometry versus OCP relationship.

The specific contributions are threefold. First, we report, to our knowledge for the first time, an MSMR parameter set for an NCA cathode, fitted on both the lithiation and delithiation branches and averaged in the hypothesis of a purely dynamical hysteresis. Second, through a peak analysis of the differential-capacity curves at different aging states, we show that the cathode thermodynamic signature is essentially unchanged by aging, providing direct support, for NCA, to the common practice of keeping the gallery parameters constant across aging and attributing the whole-cell OCV variations to electrode-level capacity and alignment changes, in the same spirit as the NMC results of Schmitt et al.~\cite{Schmitt2021}. Third, we improve the robustness of the fit by constraining each $U^0_j$ to lie within $\pm 30$~mV of an initial guess read directly from the differential capacity peaks and by enforcing site conservation. This anchors every gallery to an observed feature, which resolves the degeneracy between broad, nearly flat galleries and the accessible-stoichiometry window, and it is what gives the fitted parameters their interpretation: only a gallery that corresponds to a phase transition can be expected to survive aging unchanged, as the analysis of Section~\ref{sect:results} assumes. In addition, we use an improved, scale-invariant cost function to prevent the optimizer from collapsing the accessible-stoichiometry window. Viewed as a whole, this work brings together in a single framework the MSMR description of Verbrugge and coworkers~\cite{Verbrugge2016, Verbrugge2017}, the half-cell processing pipeline of Lu et al.~\cite{Lu2021} and the aging-invariance question investigated by Schmitt et al.~\cite{Schmitt2021}, and extends their combination to a material's chemistry and to aging states that none of them covered. The analysis here is restricted to the NCA cathode, since the treatment of a silicon-graphite anode involves a higher number of galleries and brings about the additional complexity of separating the contributions from the two active materials to the electrode OCP. We leave the parametrization of the anode material to a future work.

The remainder of the paper is organized as follows. Section~\ref{sect:experimental} describes the cycle-aging campaign, cell teardown and coin-cell fabrication, and the slow-cycling protocol. Section~\ref{sect:methods} recalls the MSMR model, details our optimization procedure and presents the cost function. Section~\ref{sect:results} discusses the extracted OCP curves, the histogram smoothing, the differential-capacity peak analysis across aging states, and the fitted and averaged MSMR parameters. Section~\ref{sect:discussion} places the contributions in the context of previous work and discusses the methodological choices, the limitations of the study and the outlook. Finally, Section~\ref{sect:conclusions} presents the conclusions.

\section{Experimental}\label{sect:experimental}

The cells belong to a larger cycle-aging campaign on commercial Molicel INR-18650-M35A and LG INR-21700-M50LT, whose test procedure and results up to 260 equivalent full cycles (EFC) were reported in an earlier work~\cite{Galasso2024}. For the present study, cycling of the Molicel cells was continued beyond that point, with the same alternation of checkup (CU) and cycling phases. The aspects of the procedure relevant to this work are summarized below.

Eight Molicel cells were tested: two were cycled over a full-SOC window (0--\SI{90}{\percent}), two over a low-SOC window (0--\SI{30}{\percent}), two over a middle-SOC window (30--\SI{60}{\percent}) and two over a high-SOC window (60--\SI{90}{\percent}). The battery tester Neware BTS4000-5V12A was used for testing the cells, while the ambient temperature was held at \SI{25}{\celsius} using a Memmert IPP750+ chamber. The cycling current was set to C/2 for both charge and discharge. Here and in the following, the C-rate expresses the current relative to the capacity of the cell under test: a current of 1\,C transfers the full cell capacity in \SI{1}{\hour}, so that, for example, C/2 completes a charge or discharge in \SI{2}{\hour}. Testing two nominally identical cells per condition allowed us to quantify the cell-to-cell variation of the capacity fade and of the OCV changes along a given aging path. Cycling over different SOC windows, in turn, was intended to activate different degradation mechanisms at the active-material level~\cite{Chowdhury2024, Kirkaldy2024}; these generally lead to different changes in the full-cell OCV and in the electrode OCP curves, even when two aging paths are compared at the same SOH~\cite{Dubarry2012, Schmitt2021}.

\begin{figure}
	\centering
	\includegraphics[width=0.6\textwidth]{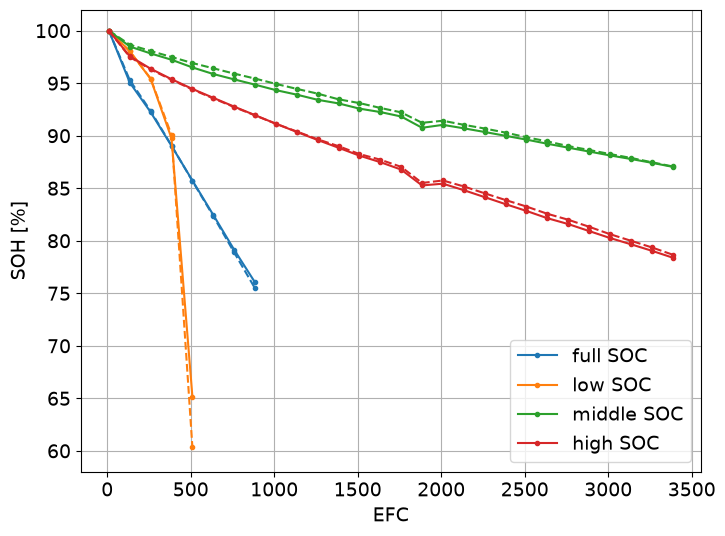}
	\caption{Aging status of the eight Molicel INR-18650-M35A undergoing cycle aging. For each cycling window, each cell is displayed with a different line style (solid and dashed). Cells aged at full-SOC and low-SOC conditions reached the EOL. The capacity oscillation observed at about 1800 EFC is due to a malfunctioning of the temperature chamber.}
	\label{fig:agingstatus}
\end{figure}

The residual capacity was measured during a CU phase carried out every 120 EFC of cycle-aging. The CU phase also included a C/30 charge and discharge, used to measure the quasi-OCV (qOCV) curve, and a dynamic part based on the worldwide harmonized light vehicles test procedure (WLTP)~\cite{UNECE2014}, which is not used in this work. Since the CU phase takes 5 EFC and each cell underwent about 10 EFC of initial characterization, CUs fall at $10 + 125n$ EFC, with $n \in \mathbb{N}$. The two pairs of cells cycled over the full- and the low-SOC windows reached their EOL state, which we define as the point at which the CU can no longer be completed, because the cell capacity and internal resistance have changed so much that the dynamic part can no longer be run without violating the safety voltage limits. This typically corresponds to a SOH below \SI{80}{\percent}, as shown in Fig.~\ref{fig:agingstatus}. Apart from the EOL point of the cells cycled at low SOC, the cell-to-cell differences in SOH remain below \SI{1}{\percent} and can therefore be neglected. The qOCV curves of the two pairs of cells that reached EOL also remain nearly identical across aging, with cell-to-cell differences an order of magnitude below the aging-induced changes (Section~S1 of the Supplementary Information). In order to investigate the aging invariance of the cathode at the deepest aging states available, one cell from each of the two pairs aged in the full- and the low-SOC windows was disassembled in the present work to fabricate half-cells.

A total of three cylindrical cells were disassembled: a pristine one, one cell cycle-aged in the full-SOC window (EOL after 885 EFC, at \SI{76}{\percent} SOH) and one cell cycle-aged in the low-SOC window (EOL after 510 EFC, at \SI{65}{\percent} SOH). Before disassembly, each cell was discharged with a constant current of C/2 to \SI{2.5}{\volt} and then held at \SI{2.5}{\volt} until the current dropped below C/100, as done in a previous study~\cite{Stapf2025}. Fully discharging the cells in this way minimizes their residual energy and lowers the safety risk during opening~\cite{Waldmann2016}. The pristine cell was opened in a nitrogen-filled glovebox (VGB-2Y, MTI Corporation), whereas the aged cells were opened in an argon-filled glovebox (GS Mega E-Line). The cell can was cut with a pipe cutter along the groove running around the cell near the cathode terminal, after which it was peeled back with pliers and the jelly roll was unrolled. Rectangular samples were cut from the electrode sheets. Since the current collector is coated with active material on both sides, whereas a single-coated electrode is required to build the coin cells, the active material was removed mechanically from one side of the electrode sheet by scraping with cotton swabs wetted with water (for the pristine cell) or dimethyl carbonate (for the aged cells). Circular samples \SI{9.5}{\milli\metre} in diameter were then punched from the single-coated sheets and transferred out of the glovebox.

The half-cells were assembled as CR2032 coin cells in an argon-filled glovebox (Jacomex). A lithium metal disk (thickness \SI{0.6}{\milli\metre}, \SI{99.9}{\percent}, Merck) served as the counter and reference electrode, and the prepared sample as the working electrode. A Whatman glass fiber separator (borosilicate glass microfiber, nominal particle retention \SI{1.5}{\micro\metre}) was placed between the electrodes and soaked with an electrolyte consisting of 1 M \ch{LiPF6} in EC/DMC (1:1, v/v). After assembly, the coin cells were left at open-circuit conditions for \SI{12}{\hour} to let the electrolyte soak into the electrodes. We manufactured a total of nine coin cells, three for each aging state of the cathode, which we marked as B1, B2, B3 for the BOL condition, F1, F2, F3 for those whose active material was aged under full-SOC cycling, and L1, L2, L3 for those whose active material was aged under low-SOC cycling. Cell L3 displayed zero voltage after manufacturing (short-circuited), while the other eight cells underwent low-current cycling to measure the OCP.

The cycling current was derived from the mass of active material and the specific capacity of NCA. Before assembling the coin cells B1, B2 and B3, each electrode disk was weighed after the coating had been removed from one side, using a precision balance (Ohaus PR224). As a reference, an identical disk was weighed with the coating still present on both sides. The active-material mass of each single-coated disk was then obtained as the difference between the double-coated and single-coated masses (Table~\ref{tab:circularsamples}). Rounding the active-material mass up to \SI{18}{\milli\gram} and assuming a typical accessible NCA capacity of \SI{200}{{mAh}\per\gram} in commercial cells~\cite{Nitta2015}, we obtain a coin-cell capacity of \SI{3.6}{{mAh}}. Aiming at a rate of C/100, we therefore cycled the coin cells at a constant current of \SI{36}{\micro\ampere}. The cathode half-cells were cycled between 3.0 and \SI{4.3}{\volt} for four full cycles, each consisting of a constant-current (CC) charge to \SI{4.3}{\volt} followed by a CC discharge to \SI{3.0}{\volt}. A \SI{6}{\hour} rest was inserted after every charge and discharge step to allow the cell voltage to relax. Cycling was performed with a Neware BTS4000-5V50mA battery tester, while the temperature was maintained at \SI{25}{\celsius} using a Neware MHW-25-S-16CH thermal chamber.

\begin{table}
	\centering
	\begin{tabular}{cccc}
	\toprule
	Coin & Mass & Mass & Mass \\
	cell & (double-coated) [mg] & (single-coated) [mg] & (active material) [mg] \\
	\midrule
	B1 & \multirow{3}{*}{37.5} & 20.2 & 17.3 \\
	B2 &                       & 19.9 & 17.6 \\
	B3 &                       & 20.3 & 17.2 \\
	\bottomrule
	\end{tabular}
	\caption{Masses of the three circular cathode disks punched from the pristine cell, from which the coin cells B1, B2 and B3 were built. A single representative double-coated disk was also weighed; its mass is therefore common to all rows. The active-material mass of each disk follows as the difference between the double-coated and the single-coated masses.}
	\label{tab:circularsamples}
\end{table}

\section{Methods}\label{sect:methods}

Because we cycle each half cell directly between the two voltage cut-offs at \SI{36}{\micro\ampere}, we assign a relative degree of lithiation $\tilde{\theta} = 0$ to the state reached at the upper cut-off $v_\mathrm{max} = \SI{4.3}{\volt}$, where the cathode is most delithiated, and $\tilde{\theta} = 1$ to the state reached at the lower cut-off $v_\mathrm{min} = \SI{3.0}{\volt}$, where it is most lithiated. In the half-cell, the charge extracts lithium from the cathode (delithiation) and the discharge inserts it (lithiation). With this definition each branch spans the whole interval $\tilde{\theta} \in [0,1]$ by construction, so we do not encounter the ``missing-data'' problem of Lu et al.~\cite{Lu2021}, that is the unobservability, for certain calibration protocols, of the portion of the discharge curve near $\tilde{\theta}=1$ and of the charge curve near $\tilde{\theta}=0$. We are therefore left only with the ``data-quality'' problem (the spurious peaks and oscillations that voltage quantization and measurement noise introduce in the differential capacity when the raw voltage is numerically differentiated) and the ``inaccessible-lithium'' problem (the fact that, owing to the finite cut-off voltages, the electrode is neither fully delithiated at $\tilde{\theta}=0$ nor fully lithiated at $\tilde{\theta}=1$).

\begin{figure}
	\centering
	\includegraphics[width=0.6\textwidth]{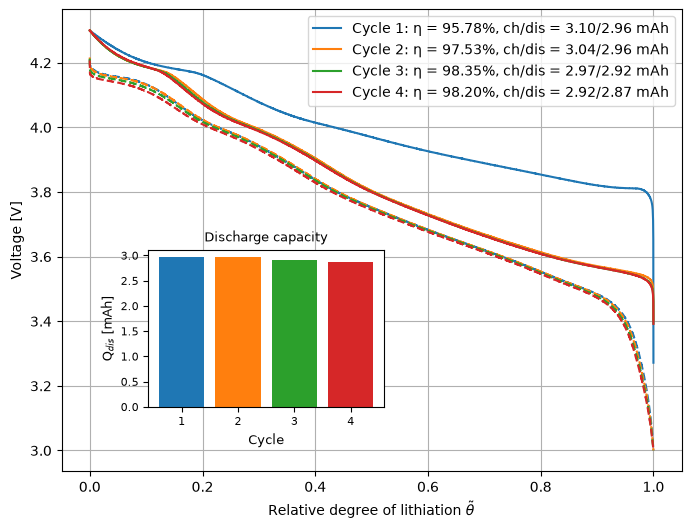}
	\caption{Measured voltage versus computed relative degree of lithiation for the four cycles of cell B2, one of the coin cells fabricated from pristine cathode material. Solid lines denote the charge (delithiation) branches and dashed lines the discharge (lithiation) branches. The first charge curve is a clear outlier, attributed to the formation processes occurring during the initial cycles, after which the Coulombic efficiency progressively stabilizes. The inset bars show the discharge capacity of each~cycle.}
	\label{fig:rawdata}
\end{figure}

The raw charge and discharge OCP curves were taken from the measured voltage during the fourth and last cycle. Voltage was sampled at \SI{1}{\second} intervals, while the capacity was obtained by integrating the constant current. Figure~\ref{fig:rawdata} plots the measured voltage against the computed relative degree of lithiation for the four cycles of cell B2: the first charge curve is a clear outlier and the Coulombic efficiency stabilizes progressively over the subsequent cycles. We attribute this behavior to the formation processes taking place during the initial cycles and therefore extract the OCP from the fourth cycle, which is the most stable. Each OCP curve used in this work is thus a single half-cycle of a single coin cell. We did not average over replicates, because small voltage shifts between cells would broaden the differential-capacity peaks on which the MSMR parametrization relies. Among the replicates of each aging state, we selected the cell with the smallest current offset, defined as the constant current that must be subtracted from the measured one so that the last two full charges end at the same charge level. The offset is expressed as a percentage of the cycling current, and a large value signals parasitic currents in the cell. At BOL, cell B1 short-circuited shortly after the test started, and cell B3 displayed a current offset of \SI{2.26}{\percent}, against \SI{0.18}{\percent} for cell B2, which was selected. At full SOC, all three cells completed the test and cell F3 had the smallest offset (\SI{0.04}{\percent}). At low SOC, cell L1 (offset \SI{0.68}{\percent}) was the only one that completed the test, since cell L3 was short-circuited at assembly and cell L2 short-circuited just after starting the test. The three full-SOC replicates agree within \SI{8.5}{\milli\volt} RMSE on charge and \SI{13.1}{\milli\volt} RMSE on discharge (Section~S2 of the Supplementary Information).

\begin{figure}
	\centering
	\begin{minipage}{0.47\textwidth}
		\centering
		\includegraphics[width=\linewidth]{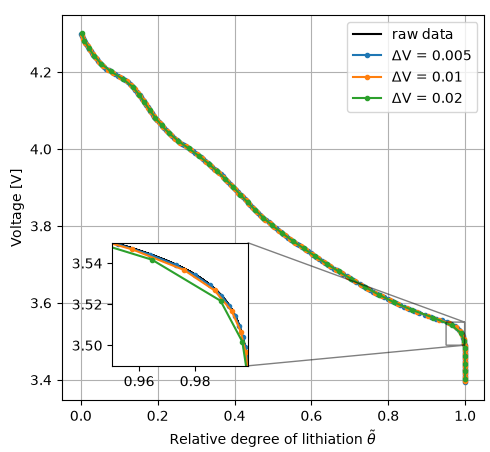}\\[4pt]
		(a) OCV\par\vspace{8pt}
	\end{minipage}
	\hfill
	\begin{minipage}{0.45\textwidth}
		\centering
		\includegraphics[width=\linewidth]{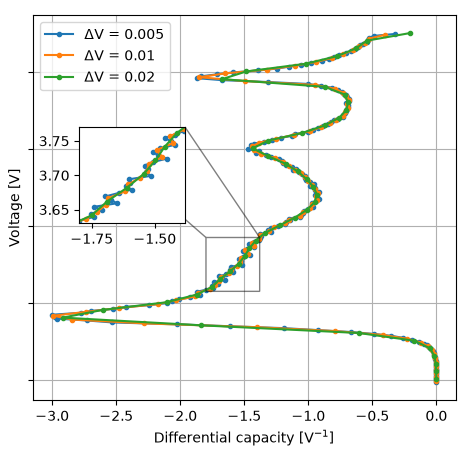}\\[4pt]
		(b) Differential capacity\par\vspace{8pt}
	\end{minipage}
	\caption{Calculation of the differential capacity with the histogram method for the BOL charge curve. The value $\Delta v = \SI{0.01}{\volt}$ provides a compromise between smoothing unphysical oscillations in the flat regions of the differential capacity and preserving the peaks.}
	\label{fig:smoothing}
\end{figure}

We address the ``data-quality'' problem with the histogram-counting method of Lu et al.~\cite{Lu2021}, which avoids the noise amplification inherent in numerically differentiating the raw voltage. The measured voltage window is divided into equally sized bins of width $\Delta v$, and the data points falling in each bin are counted. Since the voltage is sampled at a constant time interval and the current is constant, the number of samples in a bin is proportional to the charge passed while the cell traversed that voltage interval, and therefore to the differential capacity $\mathrm{d}\tilde{\theta}/\mathrm{d}\tilde{U}_\mathrm{ocp}$. A smoothed OCP curve is then recovered by integrating the smoothed differential capacity. For our cathode half-cells we adopt a bin width $\Delta v = \SI{0.01}{\volt}$, which balances the smoothing of the flat regions against the retention of peaks (Fig.~\ref{fig:smoothing}). We also studied the dependence of our results on the bin width: halving it leaves the fitted parameters and the RMSE values reported in Section~\ref{sect:results} practically unchanged, whereas doubling it starts to blur the narrowest peak (Section~S5 of the Supplementary Information).

Once the ``data-quality'' problem has been solved, we are left with the ``inaccessible-lithium'' problem, which we solve by fitting the MSMR model to obtain a mathematical description of the OCP curve defined at all values of the absolute stoichiometry. The OCP of an intercalation electrode departs from ideal Nernstian behavior because lithium insertion proceeds through several structurally distinct phases, producing a sequence of plateaus in the equilibrium-potential curve. The MSMR model of Verbrugge \textit{et~al.}~\cite{Verbrugge2017} captures this behavior by representing the electrode as a set of $J$ independent
insertion reactions (or ``galleries'') operating in parallel, each described by a thermodynamic relation between its equilibrium potential and its fractional occupancy. For gallery $j$,
\begin{equation}
  U_j = U_j^{0} + \frac{\omega_j}{f}\,
  \ln\!\left(\frac{X_j - x_j}{x_j}\right),
  \qquad f = \frac{F}{RT},
  \label{eq:msmr_gallery}
\end{equation}
where $U_j^{0}$ is the standard potential of the reaction, $\omega_j$ a dimensionless disorder parameter ($\omega_j\!\to\!0$ yields a step-like, ideal two-phase plateau, $\omega_j\!=\!1$ the single-phase Nernstian limit, and $\omega_j\!>\!1$ a disordered, solid-solution insertion), and $X_j$ the fraction of the total host sites belonging to gallery $j$ (so that $\sum_{j=1}^{J} X_j = 1$). The variable $x_j$ is the occupancy of gallery $j$ expressed as a fraction of the total host sites, so that $0 \le x_j \le X_j$ and the electrode stoichiometry is $\theta = \sum_{j=1}^{J} x_j$. Here $F$, $R$ and $T$ are Faraday's constant, the gas constant and the absolute temperature, respectively. At equilibrium all galleries share the common electrode potential, $U_\mathrm{ocp} = U_j$ for all $j$.

Inverting~\eqref{eq:msmr_gallery} for the occupancy of each gallery and summing over all galleries gives the electrode stoichiometry as an explicit function of potential,
\begin{equation}
  \theta (U_\mathrm{ocp}) = \sum_{j=1}^{J}
  \frac{X_j}{1+\exp\!\big[f\,(U_\mathrm{ocp}-U_j^{0})/\omega_j\big]},
  \label{eq:msmr_theta}
\end{equation}
whose derivative yields the differential capacity in closed form,
\begin{equation}
  \frac{\mathrm{d} \theta}{\mathrm{d} U_\mathrm{ocp}} =
  -\sum_{j=1}^{J} \frac{X_j\,f}{\omega_j}\,
  \frac{\exp\!\big[f\,(U_\mathrm{ocp}-U_j^{0})/\omega_j\big]}
       {\Big\{1+\exp\!\big[f\,(U_\mathrm{ocp}-U_j^{0})/\omega_j\big]\Big\}^{2}}.
  \label{eq:msmr_dq}
\end{equation}
Equation~\eqref{eq:msmr_dq} maps each potential plateau onto a peak in $\mathrm{d} \theta / \mathrm{d} U_\mathrm{ocp}$ centered at $U_j^{0}$, with a width governed by $\omega_j$ and an area set by $X_j$; this correspondence makes the differential-capacity curve a convenient template for initializing the MSMR model parameters.

Because the half cell is cycled only between the cut-off voltages $v_\mathrm{min}$ and $v_\mathrm{max}$, the experiment never spans the full lithiation range of the NCA electrode: it accesses only $\theta \in [\theta_\mathrm{min}, \theta_\mathrm{max}]$, where $\theta_\mathrm{min}$ and $\theta_\mathrm{max}$ are the (unknown) absolute stoichiometries reached at $v_\mathrm{max}$ and $v_\mathrm{min}$, respectively. The processed discharge/charge data therefore yield a \emph{relative} stoichiometry $\tilde{\theta} \in [0,1]$ and a corresponding relative OCP $\tilde{U}_\mathrm{ocp}(\tilde{\theta})$, related to the \emph{absolute} quantities required by a physics-based model through the transformation
\begin{align}
\theta &= \theta_\mathrm{min}
+ \tilde{\theta} \, (\theta_\mathrm{max}-\theta_\mathrm{min})
\label{eq:rel2abs}\\[1.5ex]
U_\mathrm{ocp}(\theta) &=
\tilde{U}_\mathrm{ocp}(\tilde{\theta})
\label{eq:rel2abs_U}\\
\frac{\mathrm{d}\theta}{\mathrm{d}U_\mathrm{ocp}(\theta)} &=
(\theta_\mathrm{max} - \theta_\mathrm{min})\,
\frac{\mathrm{d}\tilde{\theta}}{\mathrm{d}\tilde{U}_\mathrm{ocp}(\tilde{\theta})}
\label{eq:rel2abs_dthetadU}
\end{align}
The potential itself is unchanged by the transformation (only the abscissa is rescaled) while the differential capacity is scaled by the constant factor $(\theta_\mathrm{max}-\theta_\mathrm{min})$. Recovering $\theta_\mathrm{min}$ and $\theta_\mathrm{max}$ is thus equivalent to resolving the inaccessible-lithium problem, and is precisely what allows $\tilde{U}_\mathrm{ocp} (\tilde{\theta})$ to be re-expressed as the absolute $U_\mathrm{ocp} (\theta)$.

We fit the MSMR model parameters with the trust-region constrained algorithm (\texttt{trust-constr}), a member of the trust-region class of methods~\cite{Conn2000}, as implemented in SciPy~\cite{Virtanen2020}. The number of galleries $J$ is fixed beforehand to match the number of visible peaks in the measured differential capacity, and the remaining parameters are collected in the vector $\boldsymbol{p} = \{\theta_\mathrm{min},\theta_\mathrm{max}, U_j^{0},X_j,\omega_j\}_{j=1}^{J}$. Because the optimizer is sensitive to the starting point, the initial guesses are read directly from the differential-capacity curve: $U_j^{0}$ from the peak positions, $\omega_j$ from the peak widths in the plot of $\mathrm{d}\tilde{\theta} / \mathrm{d}\tilde{U}_\mathrm{ocp}$ versus $\tilde{U}_\mathrm{ocp}$, and $X_j$ from the peak extents in the plot of $\mathrm{d}\tilde{\theta} / \mathrm{d}\tilde{U}_\mathrm{ocp}$ versus $\tilde{\theta}$. At each iteration, the current estimates of $\theta_\mathrm{min}$ and $\theta_\mathrm{max}$ convert the processed data to absolute coordinates through Eqs.~\eqref{eq:rel2abs}--\eqref{eq:rel2abs_dthetadU}, yielding the target stoichiometry $\theta_n^{\mathrm{data}}$ and differential capacity $(\mathrm{d}\theta/\mathrm{d}U_\mathrm{ocp})_n^{\mathrm{data}}$ at each of the $N$ sampled voltages, while~\eqref{eq:msmr_theta} and~\eqref{eq:msmr_dq} supply the corresponding model values $\theta_n^{\mathrm{model}}$ and $(\mathrm{d}\theta/\mathrm{d}U_\mathrm{ocp})_n^{\mathrm{model}}$. The parameters are then identified as
\begin{align}
\boldsymbol{p}^{*} ={}&
\operatorname*{arg\,min}_{\boldsymbol{p}}
\frac{1}{(\theta_{\mathrm{max}} - \theta_{\mathrm{min}})^2} \sum_{n=1}^{N} \Biggl(
\left[\theta_n^{\mathrm{data}} - \theta_n^{\mathrm{model}}\right]^{2}
\nonumber\\
&+ w\left[
\left(\frac{\mathrm{d}\theta}{\mathrm{d} U_\mathrm{ocp}}\right)_n^{\mathrm{data}}
- \left(\frac{\mathrm{d}\theta}{\mathrm{d} U_\mathrm{ocp}}\right)_n^{\mathrm{model}}
\right]^{2}
\Biggr),
\label{eq:cost}
\end{align}
where the sum runs over the $N$ sampled voltages and the scalar weight $w$ balances the OCP residual against the differential-capacity residual. As in the work by Lu et al.~\cite{Lu2021}, we use $w = \SI{0.001}{\volt^{2}}$. To enforce physical bounds and reduce the size of the search space, we constrain the optimized values of $U^0_j$ to lie within \SI{30}{\milli\volt} of their initial guesses, and we impose $\sum_j X_j = 1$ and $\omega_j \in [0.001, 6.0)$. The lower bound on $\omega_j$ avoids infinitely sharp peaks, while the upper bound prevents the fitting of excessively broad peaks, which would absorb the background noise and lose their correspondence with a phase transition in the active material. The $\pm \SI{30}{\milli\volt}$ constraint on $U^0_j$, which may seem stringent at first sight, keeps every gallery anchored to a visible peak in the differential capacity and prevents the optimizer from replacing it with a broad, featureless ``background'' gallery that lowers the cost function without corresponding to any observable peak. Such a gallery appears in the widely used graphite parametrization of Verbrugge et al.~\cite{Verbrugge2017}, where one reaction is broad enough to have no distinct counterpart in the measured differential capacity. The present constraint is designed to keep every gallery tied to a measurable feature and therefore physically interpretable.

Our cost function~\eqref{eq:cost} differs from that of Lu et al.~\cite{Lu2021} by the additional factor $1 / (\theta_\mathrm{max} - \theta_\mathrm{min})^2$. We found this modification indispensable: since both residuals in~\eqref{eq:cost} scale with $(\theta_\mathrm{max} - \theta_\mathrm{min})$ through the change of coordinates of Eqs.~\eqref{eq:rel2abs}--\eqref{eq:rel2abs_dthetadU}, the unscaled cost can be reduced artificially, on the very same experimental data, simply by shrinking the accessible span $(\theta_\mathrm{max} - \theta_\mathrm{min})$. Normalizing by $(\theta_\mathrm{max} - \theta_\mathrm{min})^2$ makes the cost invariant to this trivial rescaling and prevents the optimizer from collapsing the stoichiometry window.

We also modify ``Method 5'' of Lu et al.~\cite{Lu2021} in two ways, which make it better suited to our experimental data. First, the charge and discharge branches are fitted simultaneously rather than independently: each branch keeps its own gallery parameters $\{U_j^{0}, X_j, \omega_j\}$, but the two branches share the same $\theta_\mathrm{min}$ and $\theta_\mathrm{max}$, and the cost function is the sum of the costs~\eqref{eq:cost} of the two branches. Sharing the window is physically justified, since $\tilde{\theta} = 0$ and $\tilde{\theta} = 1$ denote the same two states of the electrode on both branches, each branch starting from the state in which the other one ended. Second, we exclude from each branch its polarized end near the cut-off voltage, i.e., the bins of the charge branch above, and those of the discharge branch below, the voltage reached at the end of the \SI{6}{\hour} rest that follows the respective cut-off. No equilibrium state within the measured window has a potential beyond these relaxed voltages, so the excluded data are dominated by the polarization that builds up as the branch approaches its cut-off. Although they cover only a few percent of $\tilde{\theta}$, they occupy many voltage bins: for cell B2, for example, the excluded part of the discharge branch spans about \SI{5}{\percent} of $\tilde{\theta}$ but 39 of its 121 bins. If kept in the cost function, these data push $\theta_\mathrm{min}$ on the charge branch and $\theta_\mathrm{max}$ on the discharge branch to their physical limits of 0 and 1. Excluding them, instead, improves the stability of the MSMR model fit and leads to physically meaningful values of $\theta_\mathrm{min}$ and $\theta_\mathrm{max}$.

\section{Results}\label{sect:results}

The determination of the MSMR model parameters proceeds in three steps. First, we analyze how the differential-capacity peaks evolve across aging states and use this analysis to fix the number of modeled reactions $J$. Second, we refine the initial guesses read from the differential capacity by hand, visually comparing the model OCP and differential capacity with the experimental curves. Finally, we run the \texttt{trust-constr} optimizer to obtain the converged MSMR parameters.

\subsection{Peak analysis}

\begin{figure}
	\centering
	\includegraphics[width=\textwidth]{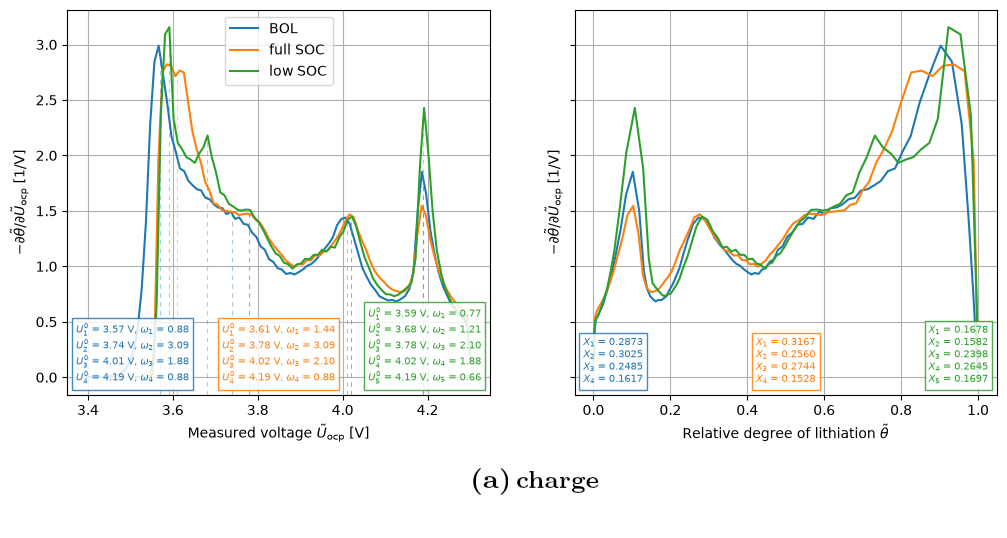}
	\includegraphics[width=\textwidth]{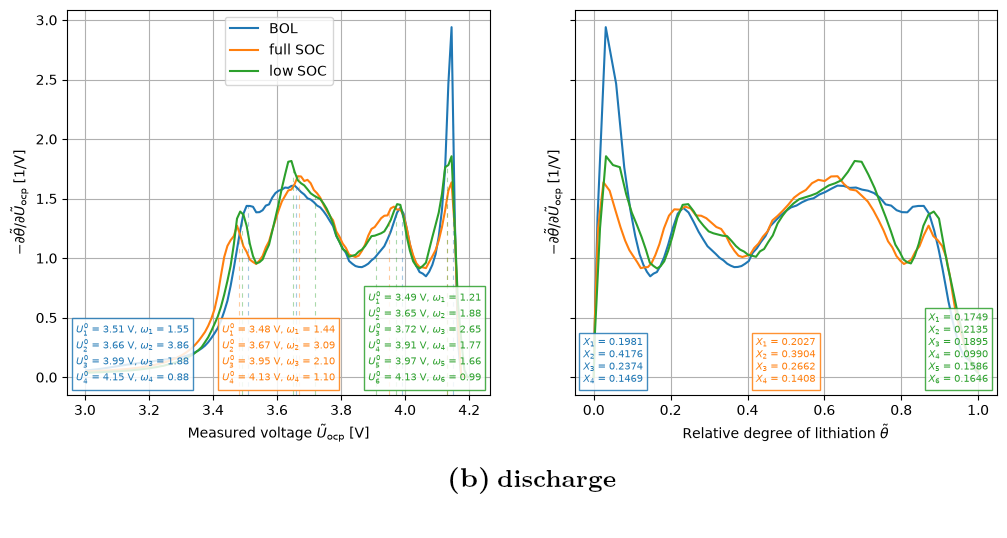}
	\caption{Differential capacity versus the experimental voltage (left) and the relative degree of lithiation (right), overlaid and color-coded by aging state (BOL, full SOC, low SOC), for (a) the charge and (b) the discharge OCP curves of the three test cells. The dashed vertical lines mark the standard potentials $U^0_j$; the color-matched boxes report the corresponding $U^0_j$ and $\omega_j$ (left) and the site fractions $X_j$ (right).}
	\label{fig:peaks}
\end{figure}

The number of reactions $J$ used to model the electrode response is chosen from the number of visible peaks in the differential-capacity plot. Figure~\ref{fig:peaks} shows the differential capacity versus the experimental voltage $\tilde{U}_\mathrm{ocp}$ (left) and versus the relative degree of lithiation $\tilde{\theta}$ (right) at the three aging states, separately for the (a) charge and (b) discharge curves. The guessed values of $U^0_j$ correspond to the peak positions in the left plots, the $X_j$ to the $\tilde{\theta}$-interval spanned by each peak in the right plots, and the $\omega_j$ are obtained from the full width at half maximum ($\mathrm{FWHM}_j$) of the peaks in the left plots, which for a single gallery is related to $\omega_j$ by
\begin{equation}
	\mathrm{FWHM}_j = \frac{4 \ln(1 + \sqrt{2})}{f} \, \omega_j .
	\label{eq:fwhm}
\end{equation}

For the charge curve, both the BOL cathode and the EOL cathode aged with full-SOC cycling ($\mathrm{SOH} = \SI{76}{\percent}$) display four peaks, whose positions $U^0_j$ shift by no more than \SI{40}{\milli\volt} from one aging state to the other, while the $X_j$ and $\omega_j$ vary by less than \SI{50}{\percent} across aging states. We stress, however, that these two parameters are only approximate, being harder to estimate manually than $U^0_j$. In the full-SOC-aged cathode (orange curve in Fig.~\ref{fig:peaks}a), peak~1 is beginning to split into two, but we assume it can still be modeled by a single peak centered between them.

The EOL cathode aged with low-SOC cycling ($\mathrm{SOH} = \SI{65}{\percent}$, green curve in Fig.~\ref{fig:peaks}a) stands apart, displaying five peaks: the splitting of peak~1 already noticed in the full-SOC-aged cathode is now much more pronounced, to the point that the new peak would require its own set of MSMR parameters. Peak~1 is the electrochemical fingerprint of the phase transition between the hexagonal H1 and the monoclinic M phases of the Ni-rich layered oxide~\cite{Nuroldayeva2023, Betzin2018}, which occurs at high degrees of lithiation, at the beginning of the charge. Its splitting indicates that the cathode has separated into two populations undergoing this transition at different potentials. The two populations may differ in structure, since Ni-rich cathodes have been shown to delithiate heterogeneously, driven by the non-uniform distribution of defects such as Li/Ni disorder and microcracks~\cite{Huang2024}, or in kinetics, since the charge-transfer resistance of aged Ni-rich cathodes rises most strongly towards the state of highest lithiation~\cite{Strehle2021}, so that particles with different surface resistance would undergo the transition at different apparent potentials. In NMC-811 harvested from cycled commercial cells, the lowest-voltage peak was likewise the most affected by aging~\cite{Schmitt2021}. The more pronounced splitting in the low-SOC-aged cathode is consistent with this cathode having been cycled, in the full cell, at high degrees of lithiation, where the $\mathrm{H}_1 \to \mathrm{M}$ transition occurs. Distinguishing between these origins would require a structural characterization of the aged cathodes, which is behind the scope of this work. Apart from this feature, the remaining peaks are essentially unchanged. Since peak~1 shows no sign of splitting in the pristine cathode, and obtaining differently aged EOL cathodes requires an expensive and time-consuming experimental effort, we model NCA with $J = 4$ at all aging states, using the MSMR model parameters fitted on BOL data. The error introduced by this choice is quantified at the end of the section.

The discharge curve, shown in Fig.~\ref{fig:peaks}b, behaves similarly. The BOL cathode and the full-SOC-aged cathode each display four distinguishable peaks, and, unlike the corresponding charge curve, the full-SOC-aged discharge curve shows no evident peak splitting. For the low-SOC-aged cathode, however, both peak~2 and peak~3 split into two distinguishable peaks, which we again attribute to the aging-induced inhomogeneity of the electrode. Apart from these two extra peaks, the remaining features do not change substantially: as for the charge curve, the $U^0_j$ shift by less than \SI{40}{\milli\volt} and the $X_j$ and $\omega_j$ by less than \SI{50}{\percent} relative to their BOL values.

\subsection{Refinement of the initial guess}

\begin{figure}
	\centering
	\begin{minipage}{0.47\textwidth}
		\centering
		\includegraphics[width=\linewidth]{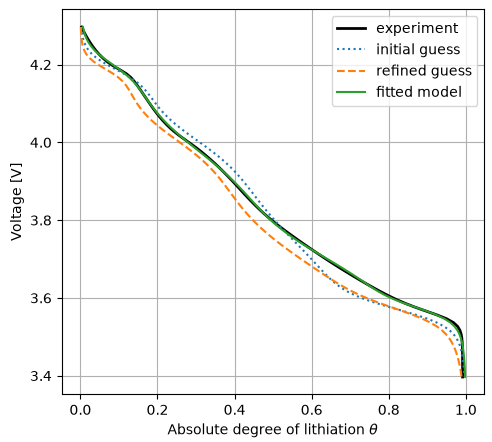}\\[4pt]
		(a) OCV\par\vspace{8pt}
	\end{minipage}
	\hfill
	\begin{minipage}{0.45\textwidth}
		\centering
		\includegraphics[width=\linewidth]{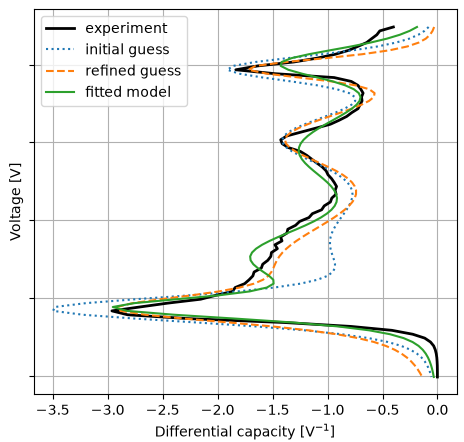}\\[4pt]
		(b) Differential capacity\par\vspace{8pt}
	\end{minipage}
	\caption{Progression of the MSMR fit for the charge curve of the BOL cathode, shown as voltage versus absolute degree of lithiation (a) and versus differential capacity (b). The experimental data (black) are overlaid with the model at three stages: the initial guess read from the peak analysis (dotted), the manually refined guess (dashed) and the optimized fit (solid).}
	\label{fig:bol-charge-fitting}
\end{figure}

Before launching the optimizer, the initial guess of MSMR model parameters from the peak analysis is refined manually: the parameters are adjusted slightly by hand while visually comparing the model curves with the experimental ones. Figure~\ref{fig:bol-charge-fitting} illustrates this step for the charge curve of the BOL cell, overlaying the raw guess from the peak analysis, the manually refined guess and the optimized fit on the experimental data. The adjustment deliberately targets the differential capacity, where the individual electrochemical reactions are most clearly resolved: we improve its agreement with the measured curve even at the cost of a small increase in the distance between the model and experimental OCP curves. Both guesses are reported in Table~\ref{tab:guesses}, so that the manual step can be inspected and reproduced. The refinement leaves the standard potentials almost untouched: on discharge every $U^0_j$ is unchanged, and on charge only $U^0_2$ is moved, by \SI{50}{\milli\volt}. The adjustment falls almost entirely on the site fractions, with a fraction of about 0.1 transferred from the first to the second gallery on both branches, and on $\omega_4$.

\begin{table}
    \centering
    \begin{tabular}{ccccc}
        \toprule
        & \multicolumn{2}{c}{Charge} & \multicolumn{2}{c}{Discharge} \\
        \cmidrule(lr){2-3} \cmidrule(lr){4-5}
        Parameter & Initial & Refined & Initial & Refined \\
        \midrule
        $U^0_1$ & 3.57 & 3.57 & 3.51 & 3.51 \\
        $U^0_2$ & 3.74 & 3.69 & 3.66 & 3.66 \\
        $U^0_3$ & 4.01 & 4.01 & 3.99 & 3.99 \\
        $U^0_4$ & 4.19 & 4.19 & 4.15 & 4.15 \\
        \midrule
        $X_1$ & 0.2873 & 0.1873 & 0.1981 & 0.0981 \\
        $X_2$ & 0.3025 & 0.4525 & 0.4176 & 0.5176 \\
        $X_3$ & 0.2485 & 0.2485 & 0.2374 & 0.2374 \\
        $X_4$ & 0.1617 & 0.1117 & 0.1469 & 0.1469 \\
        \midrule
        $\omega_1$ & 0.88 & 0.88 & 1.55 & 1.55 \\
        $\omega_2$ & 3.09 & 3.09 & 3.86 & 3.86 \\
        $\omega_3$ & 1.88 & 1.88 & 1.88 & 1.88 \\
        $\omega_4$ & 0.88 & 0.68 & 0.88 & 0.68 \\
        \bottomrule
    \end{tabular}
    \caption{Initial guess of the MSMR model parameters for the BOL cathode, read directly from the differential-capacity peaks of Fig.~\ref{fig:peaks}, and the manually refined guess from which the optimizer is launched. The standard potentials $U^0_j$ are given in volts; the site fractions $X_j$ and disorder factors $\omega_j$ are dimensionless. In both guesses $\theta_\mathrm{min}$ and $\theta_\mathrm{max}$ are initialized at 0.03 and 0.99, respectively.}
    \label{tab:guesses}
\end{table}

We refine the guesses by hand because parameters read directly from the differential-capacity plots are only approximate. Overlapping peaks make $X_j$ and $\omega_j$ hard to estimate graphically, as noted above, and even a standard potential can be misread when its peak shows up only as a shoulder of a larger one. For example, the peak of the broad second gallery of the charge branch was read about \SI{40}{\milli\volt} above its fitted $U^0_2$. Launched from the unrefined guess, the fit then stops at the edge of the \SI{30}{\milli\volt} band of this gallery. The refinement brings every guess close enough to its peak for a narrow band to contain the optimum. For NCA, whose peaks are broad and well separated, a wider band can replace the refinement: a band of \SI{50}{\milli\volt} around the unrefined guess returns the same parameters, and the fits of the aged cathodes, which are not refined by hand, use a band of \SI{100}{\milli\volt} (Sections~S3 and~S4 of the Supplementary Information). Wide bands, however, are not a general solution. In graphite, for instance, some standard potentials lie only about \SI{15}{\milli\volt} apart~\cite{Verbrugge2017}: to keep each gallery on its own peak, the band must be narrower than this spacing, and so must the error of the guess it is centered on. The refinement, therefore, does not select the solution, but it allows the band to stay narrow enough to keep every gallery anchored to its own peak, also in materials whose peaks are closely spaced.

\subsection{MSMR model optimization}

The fitted MSMR model parameters are reported in Table~\ref{tab:fitted-parameters}, and the agreement between the model and the experimental curves is shown by the optimized (solid) curve in Fig.~\ref{fig:bol-charge-fitting}. The residual mismatch of the differential-capacity curve reflects the simplicity of the MSMR model: each peak is a symmetric, bell-shaped function governed by only three parameters, whereas the real peaks are asymmetric and can depart appreciably from this shape. Despite this, the two OCP curves are almost indistinguishable. None of the inequality bounds on the MSMR parameters were active at the optimum, confirming that they acted purely as guardrails, preventing the optimizer from collapsing into unphysical regions of lower cost without otherwise constraining the solution. Moreover, the optimum does not depend on the starting point: from 50 starting points drawn at random inside the constraint box, the optimizer always returned the parameters of Table~\ref{tab:fitted-parameters}, to within \SI{0.2}{\milli\volt} in $U^0_j$, 0.0011 in $X_j$ and \SI{0.7}{\percent} in $\omega_j$ (Section~S3 of the Supplementary Information).

\begin{table}
    \centering
    \begin{tabular}{cccc}
        \toprule
        Parameter & Charge & Discharge & Average \\
        \midrule
        $\theta_\mathrm{min}$ & 0.01356 & 0.01356 & 0.01356 \\
        $\theta_\mathrm{max}$ & 0.99267 & 0.99267 & 0.99267 \\
        \midrule
        $U^0_1$ & 3.57301 & 3.50708 & 3.54004 \\
        $U^0_2$ & 3.69976 & 3.65048 & 3.67512 \\
        $U^0_3$ & 3.98634 & 4.00149 & 3.99392 \\
        $U^0_4$ & 4.20645 & 4.13729 & 4.17187 \\
        \midrule
        $X_1$ & 0.16820 & 0.02378 & 0.09599 \\
        $X_2$ & 0.36046 & 0.66711 & 0.51378 \\
        $X_3$ & 0.33618 & 0.22106 & 0.27862 \\
        $X_4$ & 0.13516 & 0.08805 & 0.11161 \\
        \midrule
        $\omega_1$ & 0.69296 & 0.56596 & 0.62946 \\
        $\omega_2$ & 2.18488 & 4.12582 & 3.15535 \\
        $\omega_3$ & 2.72906 & 2.06587 & 2.39746 \\
        $\omega_4$ & 0.87570 & 0.35273 & 0.61421 \\
        \bottomrule
    \end{tabular}
    \caption{MSMR model parameters fitted to the charge and discharge branches of the BOL cathode, and their gallery-by-gallery average. The stoichiometry window $\theta_\mathrm{min}$, $\theta_\mathrm{max}$ is shared by the two branches (Section~\ref{sect:methods}). The standard potentials $U^0_j$ are given in volts; the site fractions $X_j$ and disorder factors $\omega_j$ are dimensionless.}
    \label{tab:fitted-parameters}
\end{table}

\begin{figure}
	\centering
	\begin{minipage}{0.46\textwidth}
		\centering
		\includegraphics[width=\linewidth]{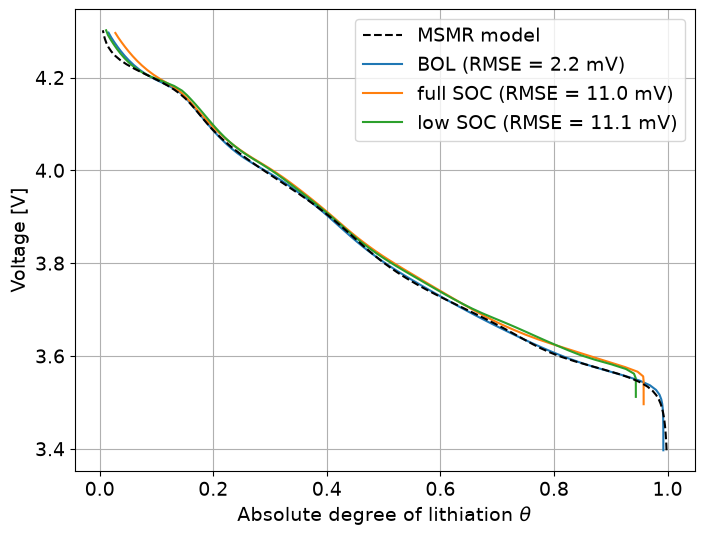}\\[4pt]
		(a) charge\par\vspace{8pt}
	\end{minipage}
	\hfill
	\begin{minipage}{0.46\textwidth}
		\centering
		\includegraphics[width=\linewidth]{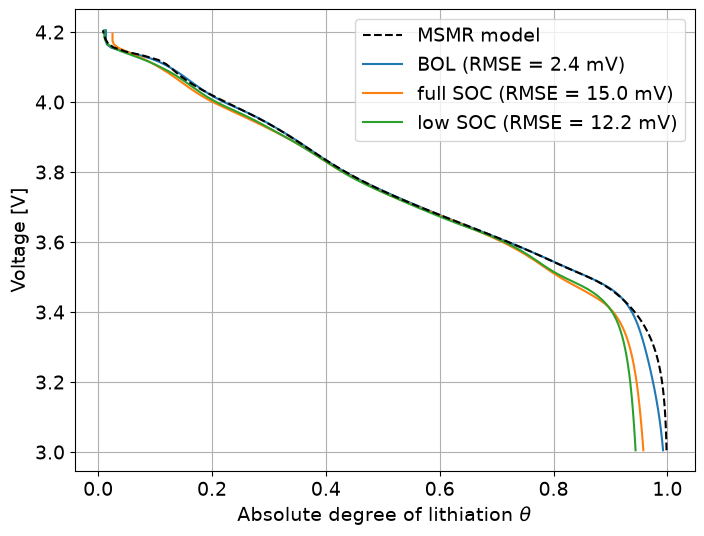}\\[4pt]
		(b) discharge\par\vspace{8pt}
	\end{minipage}
	\caption{The MSMR model fitted on the BOL cathode is compared with the experimental OCP curves of the three aging states (BOL, full SOC, low SOC), for (a) the charge curve and (b) the discharge curve. The root-mean-squared error (RMSE) of each comparison, computed over $\theta \in [0.082, 0.939]$ (without the bins beyond the relaxed voltages), is reported in the legend. The aged curves are plotted with the stoichiometry window refitted on them.}
	\label{fig:rmse}
\end{figure}

The same MSMR model fitted on the BOL cathode is compared in Fig.~\ref{fig:rmse} against the experimental OCP curves of all three aging states, separately for the charge (a) and discharge (b) branches. The root-mean-squared error (RMSE) is computed over $\theta \in [0.082, 0.939]$, similarly to Lu et al., who also evaluate the RMSE of their OCP estimates over restricted stoichiometry intervals~\cite{Lu2021}. This is the range over which both BOL branches are in steady state, from the first fitted bin of the charge branch to the last fitted bin of the discharge branch. For each branch, it leaves out both the polarized end excluded from the fit (Section~\ref{sect:methods}) and the beginning of the branch, where the polarization is still building up after the rest: there the measured voltage is still close to the OCP, whereas the fitted branch describes the OCP shifted by the fully developed polarization. For the aged cathodes, the bins beyond their own relaxed voltages are left out as well. This interval excludes the steep tails of the OCP curve, where a small stoichiometry error translates into a large voltage error. The RMSE does not quantify the accuracy of the model at the extremes of the OCP curve, including lithiation states that can in principle be reached during full-cell operation. For the aged cathodes, before computing the RMSE, the stoichiometry window $\theta_\mathrm{min}$, $\theta_\mathrm{max}$ of the BOL model is refitted on their charge and discharge curves, with all the galleries held fixed. A new window does not change the shape of the OCP curve, but only the part of it that the C/100 test covers, which depends on where the aged half-cell reaches its cut-off voltages; in a full-cell model, the capacity and alignment parameters of the electrode would absorb this change~\cite{Schmitt2021, Dubarry2012, Hu2022, Kirst2024}. The refitted windows are $[0.025, 0.958]$ for the full-SOC-aged and $[0.011, 0.944]$ for the low-SOC-aged cathode, against $[0.014, 0.993]$ at BOL. As expected, the BOL model reproduces the BOL data almost exactly (RMSE is \SI{2.2}{\milli\volt} on charge and \SI{2.4}{\milli\volt} on discharge). More notably, the same BOL galleries still describe the aged cathodes well: the RMSE grows only to \SI{11.0}{\milli\volt} on charge and \SI{15.0}{\milli\volt} on discharge for the full-SOC-aged cathode ($\mathrm{SOH}=\SI{76}{\percent}$) and to \SI{11.1}{\milli\volt} and \SI{12.2}{\milli\volt} for the more strongly degraded low-SOC-aged cathode ($\mathrm{SOH}=\SI{65}{\percent}$).

These values should be read against the cell-to-cell reproducibility of the half-cell measurement: after the same alignment of their stoichiometry axes, the three full-SOC replicates differ by at most \SI{5.3}{\milli\volt} RMSE on charge and \SI{6.0}{\milli\volt} RMSE on discharge over the same range (Section~S2 of the Supplementary Information). The aging-induced deviation thus exceeds the reproducibility by a factor of 2.0 to 2.5 on both branches. This residual bundles together the genuine change of the cathode OCP upon aging and the error of forcing the four BOL galleries onto cathodes that have developed additional peaks. Fitting the aged cathodes directly separates the two contributions (Section~S4 of the Supplementary Information): refitting the four galleries reduces the RMSE of the low-SOC-aged cathode to \SI{3.1}{\milli\volt} on charge and \SI{1.1}{\milli\volt} on discharge, and adding one gallery for each additional peak reduces it only further to \SI{0.9}{\milli\volt} and \SI{0.8}{\milli\volt}. The imposed four-gallery structure, therefore, accounts for only about \SI{2}{\milli\volt}, with nearly all of the residual stemming from aging-induced changes in the parameters of the four galleries. Its modest magnitude, together with the limited shift of the peak positions discussed in the peak analysis, indicates that the thermodynamic signature of the NCA cathode\,---\,the shape of its OCP curve\,---\,is largely preserved upon aging. This supports, for NCA, the common modeling practice of holding the intrinsic gallery parameters fixed at their BOL values and ascribing the whole-cell OCV changes to electrode-level capacity and alignment shifts~\cite{Dubarry2012}, mirroring the analogous result reported for NMC-811 by Schmitt et al.~\cite{Schmitt2021}.

Generally speaking, the aging invariance of the OCP shape is not expected for every aging mechanism, but only for those mechanisms that cause loss of active material and resistance increase without altering the remaining active material, like particle cracking and contact loss. Mechanisms that alter the bulk structure of the material, such as Li/Ni mixing, bulk phase transformations or oxygen vacancies, change the thermodynamics of lithium (de)intercalation and hence the OCP. In our cathodes, the only clear change is the peak splitting, which suggests that the active material is separating into multiple populations that undergo the same phase transitions at slightly different potentials. Apart from this splitting, the bulk structure of the cathode appears largely preserved. Consistent with our results, Strehle et al.~\cite{Strehle2021} found the layered bulk structure of another layered oxide, NMC-811, to be well preserved after 1000 cycles at room temperature.

\begin{figure}
	\centering
	\includegraphics[width=0.5\textwidth]{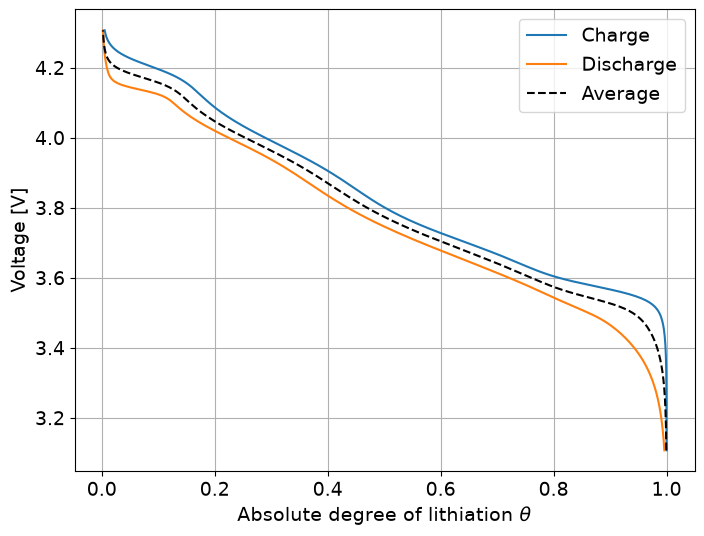}
	\caption{The OCP curve of NCA, obtained by averaging the charge and discharge MSMR model parameters of the BOL cathode.}
	\label{fig:charge-discharge-average}
\end{figure}

A single OCP curve is obtained for NCA by averaging, gallery by gallery, the MSMR parameters fitted to the charge and discharge branches (Table~\ref{tab:fitted-parameters} and Fig.~\ref{fig:charge-discharge-average}). This corresponds to ``Method 5'' of Lu et al.~\cite{Lu2021} and rests on the assumption that the charge--discharge gap stems entirely from kinetic polarization of equal magnitude on the two branches, a purely dynamic hysteresis, so that the equilibrium OCP lies midway between them. GITT measurements on commercial cells with layered-oxide positive electrodes support this assumption: they found the hysteresis of the positive electrode to be small, with the hysteresis of the full cell arising mostly from the negative electrode~\cite{Birkl2015, Chen2020}. For an Al-doped NMC-811 electrode, the GITT charge and discharge curves were found to lie much closer to each other than those measured by slow galvanostatic cycling at C/20~\cite{Chen2020}, indicating that the gap between the latter comes mostly from polarization.

\section{Discussion}\label{sect:discussion}

It is useful to state explicitly what the present work adds to the literature it builds upon. The MSMR model was formulated by Verbrugge and coworkers, who also published parameter sets for the standard chemistries, obtained however from largely unconstrained fits on pristine material~\cite{Verbrugge2016, Verbrugge2017}. Lu et al. later contributed a systematic pipeline for extracting MSMR parameters from half-cell data, addressing the data-quality, missing-data and inaccessible-lithium problems, and demonstrated it, again on pristine material, for graphite and NMC electrodes~\cite{Lu2021}. In parallel, and outside the MSMR framework, the teardown study of Schmitt et al. asked how the electrode OCP curves themselves change upon aging~\cite{Schmitt2021}. The present work combines these threads in a single framework: the MSMR model is fitted through the pipeline of Lu et al., strengthened by the constraints and the scale-invariant cost function of Section~\ref{sect:methods}, and applied to a chemistry absent from the MSMR literature, NCA, using electrode material harvested not only from a pristine cell, but also from cells aged to EOL along two different paths. The combination is beneficial in both directions: the processing pipeline provides the machinery needed to parametrize NCA reliably, while the aged material turns the parametrization into a quantitative, model-based test of the aging-invariance assumption on which MSMR-based aging models rest, extending to NCA, and to considerably deeper aging states, the answer that Schmitt et al. gave for NMC-811. The most aged of their cells still retained \SI{85.9}{\percent} SOH, whereas our cathodes were harvested from cells that reached EOL at \SI{76}{\percent} and \SI{65}{\percent} SOH.

Two methodological choices proved important for obtaining a physically meaningful fit, and neither is specific to NCA. Constraining each standard potential to within \SI{30}{\milli\volt} of an initial guess read from the differential-capacity peaks, together with enforcing site conservation, kept every gallery anchored to an observable feature and avoided the broad, featureless ``background'' galleries that can otherwise lower the cost function without physical justification. Enforcing site conservation is inexpensive, and it keeps the parameter set self-consistent for downstream use: since a published set is commonly taken as the starting point of a new fit, a total that departs from unity is carried along with it~\cite{Verbrugge2017, Garrick2024}. The scale-invariant cost function, in turn, removed the trivial degeneracy by which the optimizer could collapse the accessible-stoichiometry window. Together with the published parameter table, these ingredients make the present parametrization directly usable in physics-based cell models and in the state-estimation algorithms of battery management systems.

The study also has clear limitations. The analysis is confined to the cathode and rests on a small number of teardowns: one cell per aging state and only the two distinct aging paths that reached the end of life. The middle- and high-SOC cells, which had not reached the end of life at the time of the teardown, were cycled in SOC windows that may activate different degradation mechanisms~\cite{Chowdhury2024, Kirkaldy2024}, and whether the cathode OCP is preserved along these paths remains to be tested. In addition, the symmetric, three-parameter MSMR peak shape cannot fully reproduce the asymmetric experimental peaks, and the charge--discharge averaging assumes a purely dynamic hysteresis of equal magnitude on the two branches. GITT studies indicate that the hysteresis of layered-oxide positive electrodes is small~\cite{Birkl2015, Chen2020}, but neither this nor the symmetry of the polarization could be verified directly on our half-cells, for which no GITT data are available.

As for the outlook, the natural next step is the silicon-graphite negative electrode of the same cell. This is a substantially harder task: the anode response involves a larger number of reactions, and the contributions of graphite and silicon to the OCP and to the differential capacity must be disentangled. Moreover, the hysteresis of silicon is thermodynamic rather than purely dynamic, so its charge and discharge branches cannot be averaged as we did here for NCA and will need to be parametrized separately. Our cycle-aged cells are a distinctive asset for this purpose. Because silicon degrades preferentially and its capacity contribution diminishes with aging~\cite{Schmitt2021}, the comparison between pristine and aged anodes offers a practical route to separate the silicon and graphite contributions, an information unavailable to previous MSMR parametrizations, which have relied exclusively on pristine material~\cite{Verbrugge2016, Verbrugge2017, Lu2021}.

\section{Conclusions}\label{sect:conclusions}

In this work we reported the first MSMR model parametrization of an NCA positive electrode, obtained from half-cell OCP measurements on cathode material harvested from a commercial silicon-graphite/NCA cell at beginning of life and at the end of life of two distinct cycle-aging paths. The parameter set was fitted separately on the charge and discharge branches with a constrained, scale-invariant optimization that anchors every reaction to a visible differential-capacity peak, and the two branches were averaged under the assumption of a purely dynamic hysteresis (Table~\ref{tab:fitted-parameters}). Across aging states, the thermodynamic signature of the cathode proved largely preserved: the standard potentials shift by less than \SI{40}{\milli\volt}, the site fractions and disorder factors by less than \SI{50}{\percent}, and the BOL galleries reproduce the OCP of both aged cathodes, down to \SI{65}{\percent} SOH, within \SI{15}{\milli\volt} RMSE once the stoichiometry window is refitted.

These results provide direct experimental support, for NCA, to the widespread modeling practice of holding the intrinsic gallery parameters fixed across aging and attributing the whole-cell OCV changes to electrode-level capacity and alignment shifts, and they deliver a parameter set that is directly usable in physics-based cell models and battery management systems. Future work will extend the parametrization to the silicon-graphite negative electrode of the same cells, whose larger number of reactions and thermodynamic hysteresis require a separate treatment.

\section*{Declarations}

\bmhead{Funding}

This work was supported by the Grant Agency of the Czech Technical University in Prague, grant No. SGS24/136/OHK3/3T/13. Z.V.Ž. acknowledges the support of the Recovery and Resilience Plan for Slovakia under the project SUNFLOWERS No. 09I02-03-V01-00022, and of a grant from the Programme Johannes Amos Comenius of the Ministry of Education, Youth and Sports of the Czech Republic, No. CZ.02.01.01/00/22\_008/0004558 ``Advanced MUltiscaLe materials for key Enabling Technologies'' (AMULET).

\bmhead{Conflict of interest}

The authors declare no competing interests.

\bmhead{Ethics approval and consent to participate}

Not applicable.

\bmhead{Consent for publication}

Not applicable.

\bmhead{Data availability}

The fitted MSMR parameters that support the findings of this study are reported in full in Table~\ref{tab:fitted-parameters}, and the measured OCP and differential-capacity curves are presented in the figures. The underlying raw cycling datasets are not publicly available at this time.

\bmhead{Materials availability}

Not applicable.

\bmhead{Code availability}

The code implementing the constrained MSMR fit was developed in-house and is not publicly available.

\bmhead{Author contributions}

M.G.: Conceptualization, Methodology, Investigation, Formal analysis, Software, Visualization, Writing -- Original Draft, Writing -- Review \& Editing, Funding Acquisition. P.Č.: Investigation. I.H.: Investigation, Resources. S.S.: Conceptualization, Supervision. Z.V.Ž.: Supervision, Writing -- Review \& Editing, Funding Acquisition. V.K.: Conceptualization, Supervision, Funding Acquisition. All authors discussed the results, reviewed the manuscript and approved its final version.


\includepdf[pages=-]{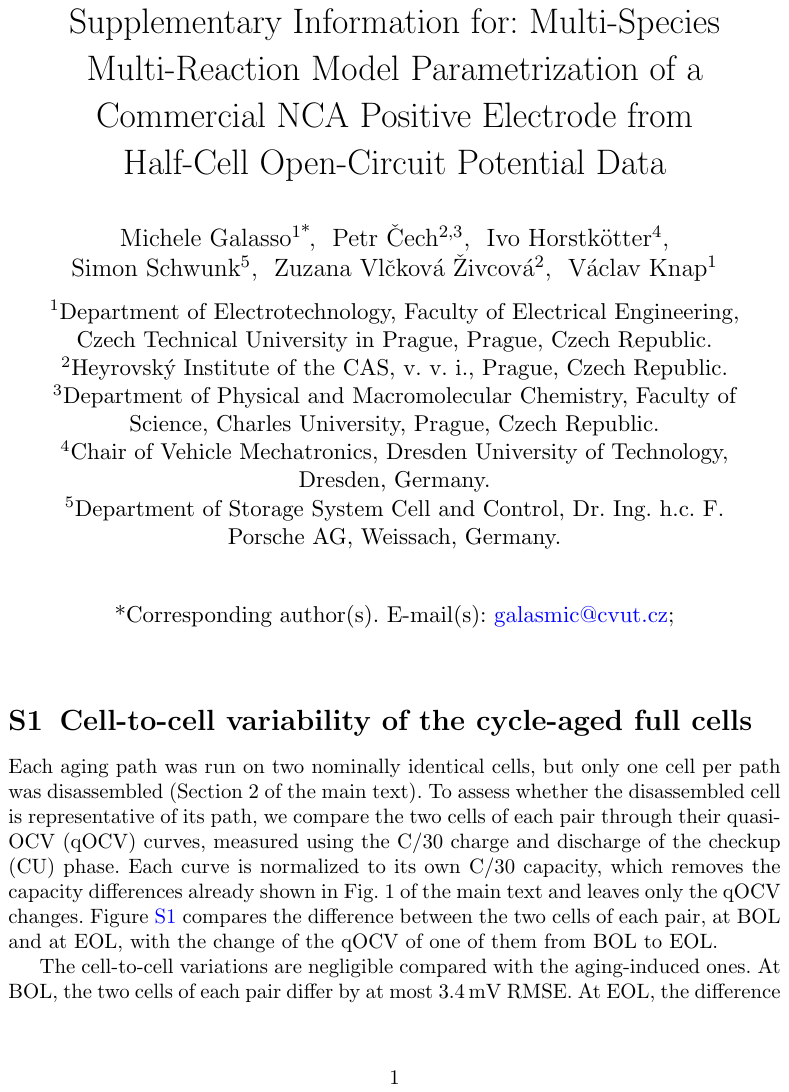}

\end{document}